\documentclass[pra,twocolumn,superscriptaddress,shownopacs,amssymb]{revtex4-2}
\usepackage{graphicx,psfrag,amsmath,amssymb}

\usepackage[usenames, dvipsnames]{color}

\usepackage{hyperref} 
\hypersetup{
    colorlinks=true,
    linkcolor=Blue,
    citecolor=Blue,
    filecolor=Blue,      
    urlcolor=Blue,
    pdftitle={Few-bosons},
    }

\begin{document}

\title{Dimers, trimers, tetramers, and other multimers in a multiband Bose-Hubbard model}

\author{M. Iskin}
\affiliation{
Department of Physics, Ko\c{c} University, Rumelifeneri Yolu, 34450 Sar\i yer, Istanbul, Turkey
}

\author{A. Kele\c{s}}
\affiliation{
Department of Physics, Middle East Technical University, Ankara, 06800, Turkey
}

\date{\today}

\begin{abstract}

We study the bound states of $N$ identical bosons that are described by a multiband 
Bose-Hubbard model with generic hoppings and an attractive onsite interaction.
Using a variational approach, we first derive exact integral equations for the dimers, 
trimers, tetramers, and other multimers, and then apply them to a one-dimensional
sawtooth model that features two bands. In particular we reveal the presence of 
not only the offsite dimer states which consist of two monomers on different 
sites even in the strong-coupling limit but also the offsite trimer states which 
consist of either a dimer on one site and a monomer on another site or 
three monomers on three different sites.
Our variational calculations for the ground states of onsite dimers, onsite trimers 
and offsite trimers benchmark perfectly well with the DMRG simulations. 
We also present DMRG results for the ground states of onsite tetramers, offsite 
tetramers, onsite pentamers, offsite pentamers, and for those of other multimers.

\end{abstract}

\maketitle

\section{Introduction}
\label{sec:intro}

Exactly solvable few-body problems~\cite{mattis86}, including but not 
limited to dimers, trimers and tetramers, in a periodic lattice have 
long been of interest to physicists in the contexts of Cooper pairs in 
superconductors, excitons and Frenkel-exciton--hole trions in semiconductors, 
multimagnons in quantum magnetism, Efimov trimers 
in BECs and other quantum gases/liquids, etc. However, since almost all of 
the studies in the literature are about single-band lattices, 
the effects of multiple bands have entirely been overlooked. 
For instance while dimers have long been known to 
be the only possible bound state between identical (i.e., equal-mass) fermions
in a single-band lattice~\cite{mattis86}, 
the energetic stability of trimers, tetramers and 
some other multimers have recently been shown in the presence of 
multiple Bloch bands~\cite{orso22, iskin22b, iskin22c}. 
It is conceivable that there may not even be an upper bound on the size of 
the possible multimers when they are formed in a flat band, albeit with
smaller and smaller binding energies~\cite{iskin22c}. 

On the other hand dimers, trimers, tetramers and all other multimers have long
been known to be allowed for identical bosons in a single-band 
lattice due to the absence of Pauli exclusion~\cite{mattis86}. 
Except for the dimers, these states form discontinuously in three dimensions 
(i.e., they have finite size at the formation 
threshold) as a function of interaction strength, so that they are already 
in the strong-coupling regime when they appear and are strongly co-localized 
on the same lattice site. 
They are sometimes called the onsite trimers, onsite tetramers, etc., 
since their binding energies are linearly proportional to the interaction 
strength in the strong-coupling limit just like those of the onsite 
dimers~\cite{valiente08, javanainen10, sanders11}. Onsite trimers form continuously 
in lower (two and one) dimensions without a threshold on the coupling strength.
Remarkably there are also weakly-bound offsite trimer states (above the ground state
of the spectrum in the strong-coupling limit) which consist of a dimer on one 
site and a monomer on another site, and whose peculiar binding mechanism turns 
out to be an effective particle-exchange interaction between the onsite dimer 
and the monomer~\cite{valiente10}. 

In this paper we study the effects of multiple Bloch bands on the formation
of few-boson bound states.
For this purpose we first derive exact integral equations for the dimers, 
trimers, tetramers, and other multimers that are described by a multiband 
Bose-Hubbard model with generic hoppings and an onsite attractive interaction.
Then, motivated primarily by our benchmarking capacity with the DMRG simulations, 
we calculate the two-body and three-body spectra in a one-dimensional sawtooth 
model that features two Bloch bands.
One of our main findings is that, in addition to the onsite dimer, onsite 
trimer and offsite trimer states, the two-band lattice also exhibits
weakly-bound offsite dimer states which consist of two monomers on 
different sites even in the strong-coupling limit. In return these offsite 
dimers also give rise to offsite trimers that consist of three monomers 
on three different sites in the strong-coupling limit. 
We show that our results for the ground states of onsite dimers, onsite 
trimers and offsite trimers perfectly benchmark with the DMRG 
simulations, and we present additional DMRG results for the ground states 
of onsite tetramers, offsite tetramers, onsite pentamers, offsite pentamers 
and for those of other multimers. 
Given that our variational results are readily applicable to all sorts of 
lattices in all dimensions, we believe they may find useful applications 
in future few-body studies. 

The rest of the paper is organized as follows. In Sec.~\ref{sec:va} we first 
introduce the Hamiltonian for the multiband Bose-Hubbard model in reciprocal 
space, and then use the variational approach to derive integral equations for 
the $N$-body bound states, including the dimers ($N = 2$), trimers ($N = 3$), 
tetramers ($N = 4$), and other multimers ($N \ge 5$). In Sec.~\ref{sec:ni}
we apply our theory to a sawtooth lattice, analyze the full dimer and trimer 
spectra, and benchmark them with the DMRG simulations. 
The paper ends with a brief summary of our findings and outlook in 
Sec.~\ref{sec:conc}.

\section{Variational Approach}
\label{sec:va}

Motivated by the success of variational approach in describing the few-fermion bound 
states in a multiband Hubbard model~\cite{iskin22c}, here we extend it to study the 
few-boson bound states in a multiband Bose-Hubbard model.

\subsection{Multiband Bose-Hubbard model}
\label{sec:mbhm}

The Bose-Hubbard model~\cite{fisher89} and its various extensions are often 
employed in the analysis of the low-temperature phases of cold bosonic atoms 
in optical lattices but not limited to them. Historically the observation of 
superfluid-Mott insulator transition was one of the most prominent achievements 
in this field. See, e.g., Refs.~\cite{greiner02, spielman08, gemelke09, bakr10}
and many others. In its simplest form, the model Hamiltonian can be written as
\begin{align}
\label{eqn:BHHam}
\mathcal{H} = -\sum_{Si; S'i'} t_{Si; S'i'} c_{S i}^\dagger c_{S' i'}
+ \frac{U}{2} \sum_{S i} 
c_{S i}^\dagger c_{S i}^\dagger 
c_{S i} c_{S i},
\end{align}
where the first term accounts for the hopping of particles from a site $S'$ in unit 
cell $i'$ to a site $S$ in unit cell $i$ with amplitude $t_{Si; S'i'}$, and the 
second term accounts for the density-density interaction between particles when 
they are on the same site. Here a positive or negative $U$ corresponds, respectively, 
to a repulsive or attractive interaction, and the prefactor $1/2$ is to avoid double 
counting. In this paper we have a generic lattice with periodic boundary 
conditions in mind, where $N_c$ is the number of unit cells in the system and $N_b$
is the number of sublattices in a given unit cell. This is in such a way that the 
total number of lattice sites in the system is $N_s = N_b N_c$.

In reciprocal space, the Bose-Hubbard model can be conveniently 
expressed as~\cite{iskin21}
\begin{align}
\label{eqn:Ham}
\mathcal{H} = \sum_{n \mathbf{k}} \varepsilon_{n\mathbf{k}}
& c_{n \mathbf{k}}^\dagger c_{n \mathbf{k}}
+ \frac{1}{2 N_c} 
\sum_{\substack{nmn'm' \\ \mathbf{k}\mathbf{k'}\mathbf{q}}}
V_{n'm'\mathbf{k'}}^{nm\mathbf{k}}(\mathbf{q})
\nonumber \\ 
&\times 
c_{n,\mathbf{k}+\frac{\mathbf{q}}{2}}^\dagger
c_{m,-\mathbf{k}+\frac{\mathbf{q}}{2}}^\dagger
c_{m',-\mathbf{k'}+\frac{\mathbf{q}}{2}}
c_{n',\mathbf{k'}+\frac{\mathbf{q}}{2}},
\end{align}
where $n$ denotes the Bloch bands, $\mathbf{k}$ is the crystal momentum in the 
first Brillouin zone (BZ), $\varepsilon_{n\mathbf{k}}$ is the corresponding 
single-particle dispersion, and
$
V_{n'm'\mathbf{k'}}^{nm\mathbf{k}}(\mathbf{q}) = U\sum_S 
n_{S, \mathbf{k}+\frac{\mathbf{q}}{2}}^*
m_{S, -\mathbf{k}+\frac{\mathbf{q}}{2}}^*
{m'}_{S, -\mathbf{k'}+\frac{\mathbf{q}}{2}}
{n'}_{S, \mathbf{k'}+\frac{\mathbf{q}}{2}}
$
characterizes the onsite interactions. This Hamiltonian simply follows from the 
Fourier expansion
$
c_{S i}^\dagger = \frac{1}{\sqrt{N_c}} \sum_\mathbf{k} 
e^{-i \mathbf{k} \cdot \mathbf{r_{S i}}} c_{S \mathbf{k}}^\dagger
$
of the site operators where $\mathbf{r_{S i}}$ is the position of the sublattice 
site $S$ in unit cell $i$, along with the basis transformation
$
c_{n \mathbf{k}}^\dagger = \sum_S n_{S \mathbf{k}} 
c_{S \mathbf{k}}^\dagger
$
from the orbital to the band basis where $n_{S \mathbf{k}}$ is the projection of the 
Bloch state onto the sublattice $S$.
Note that there are $N_b$ Bloch bands for a system that has $N_b$ sublattices in its 
unit cell, and the Bloch factors $n_{S \mathbf{k}}$ follow from the diagonalization 
of the $N_b \times N_b$ Bloch Hamiltonian. In this paper we are interested in the bound 
states of $N$ identical bosons that are described by Eq.~(\ref{eqn:Ham}).

\subsection{$N$-body bound states}
\label{sec:nbbs}

For a given center-of-mass (CoM) momentum $\mathbf{q}$, the energy $E_N^\mathbf{q}$ 
of an $N$-body bound state $|\Psi_\mathbf{q} \rangle$ follows from the Schr\"odinger 
equation
$
\mathcal{H} | \Psi_\mathbf{q} \rangle = E_N^\mathbf{q} | \Psi_\mathbf{q} \rangle.
$
In our variational approach, these bound states are described exactly by the ansatz
\begin{align}
\label{eqn:Psiq}
| \Psi_\mathbf{q} \rangle =
\sum_{\substack{n_1\cdots n_N \\ \mathbf{k_1} \cdots \mathbf{k_{N-1}}}} 
\alpha_{n_1 \cdots n_{N-1} n_N}^{\mathbf{k_1} \cdots \mathbf{k_{N-1}}} (\mathbf{q})
\bigg( \prod_{i=1}^{N} c_{n_i \mathbf{k_i}}^\dagger \bigg)
| 0 \rangle,
\end{align}
where the variational parameter 
$
\alpha_{n_1 \cdots n_{N-1} n_N}^{\mathbf{k_1} \cdots \mathbf{k_{N-1}}} (\mathbf{q})
$
is a complex number that depends on all of the band as well as momentum indices with 
the exception of $\mathbf{k_N}$. This is because 
$
\mathbf{k_N} = \mathbf{q} - \sum_{i=1}^{N-1}  \mathbf{k_i}
$
follows from the conservation of $\mathbf{q}$, and it is not an independent variable.
Note that the variational ansatz is of the most general form as long as the interaction 
term respects the translation invariance of the lattice.
The normalization condition can be written as
$
\langle \Psi_\mathbf{q} | \Psi_\mathbf{q} \rangle = 
N! \sum_{\substack{n_1 \cdots n_N \\ \mathbf{k_1} \cdots \mathbf{k_{N-1}}}} 
|\alpha_{n_1 \cdots n_{N-1} n_N}^{\mathbf{k_1} \cdots \mathbf{k_{N-1}}} (\mathbf{q})|^2,
$
where we make extensive use of the relations
\begin{align}
\alpha_{n_1 \cdots n_i \cdots n_j \cdots n_{N-1} n_N}^{\mathbf{k_1} \cdots \mathbf{k_i} 
\cdots \mathbf{k_j} \cdots \mathbf{k_{N-1}}} (\mathbf{q})
&= \alpha_{n_1 \cdots n_j \cdots n_i \cdots n_{N-1} n_N}^{\mathbf{k_1} \cdots \mathbf{k_j} 
\cdots \mathbf{k_i} \cdots \mathbf{k_{N-1}}} (\mathbf{q}),
\\
\alpha_{n_1 \cdots n_i \cdots n_{N-1} n_N}^{\mathbf{k_1} \cdots \mathbf{k_i} 
\cdots \mathbf{k_{N-1}}} (\mathbf{q})
&= \alpha_{n_1 \cdots n_N \cdots n_{N-1} n_i}^{\mathbf{k_1} \cdots \mathbf{k_N} 
\cdots \mathbf{k_{N-1}}} (\mathbf{q}),
\end{align}
that follow from the exchange symmetry of identical bosons.

After a lengthy but straightforward algebra, the expectation value of the Hamiltonian 
given in Eq.~(\ref{eqn:Ham}) can be written as
\begin{align}
&\langle \mathcal{H} \rangle = 
N! \sum_{\substack{n_1 \cdots n_{N-1} n_N \\ \mathbf{k_1} \cdots \mathbf{k_{N-1}}}} 
|\alpha_{n_1 \cdots n_{N-1} n_N}^{\mathbf{k_1} \cdots \mathbf{k_{N-1}}} (\mathbf{q})|^2 
\bigg( \sum_{i = 1}^N \varepsilon_{n_i \mathbf{k_i}} \bigg)
\nonumber \\
& + N(N-1) \frac{U}{2N_c}
\sum_{\substack{n_1 \cdots n_N m_1 m_2 \\ S \mathbf{k_1} \cdots \mathbf{k_{N-1}} \mathbf{k}}} 
\alpha_{n_1 \cdots n_{N-1} n_N}^{\mathbf{k_1} \cdots \mathbf{k_{N-1}}} (\mathbf{q})
{m_1^*}_{S \mathbf{k}} 
\\
&\times 
\sum_{P} 
\big[\alpha_{n_{i_1} \cdots n_{i_{N-2}} m_1 m_2}^{\mathbf{k_{i_1}} \cdots \mathbf{k_{i_{N-2}}} \mathbf{k}}(\mathbf{q})\big]^*
{m_2^*}_{S \mathbf{Q}}
{n_{i_{N-1}}}_{S \mathbf{k_{i_{N-1}}}} 
{n_{i_N}}_{S \mathbf{k_{i_N}}}.
\nonumber
\end{align}
Here we define
$
\mathbf{Q} = \mathbf{q} - \sum_{j=1}^{N-2}  \mathbf{k_{i_j}} - \mathbf{k}
$
for convenience, and the summation over $P$ denotes all possible permutations of 
$\{i_1, i_2, \cdots i_N \}$ subindices where each subindex refers to one of the 
$N$ values in the set $\{ 1, 2, \cdots,N \}$. This is such that 
$
\mathbf{k_{i_j}} \in \{ \mathbf{k_1}, \mathbf{k_2}, \cdots, \mathbf{k_N} \}
$
and
$
n_{i_j} \in \{ n_1, n_2, \cdots, n_N \},
$
respectively, span the corresponding momentum and band variables. 
The variational parameters are determined through the functional minimization of 
$
\langle \Psi_\mathbf{q} | \mathcal{H} - E_N^\mathbf{q} | \Psi_\mathbf{q} \rangle,
$
leading eventually to
\begin{widetext}
\begin{align}
\gamma_{n_1 \cdots n_{N-2} S}^{\mathbf{k_1} \cdots \mathbf{k_{N-2}}} (\mathbf{q})
= - \frac{N(N-1)}{N!} \frac{U}{2N_c} \sum_{n_{N-1} n_N S' \mathbf{k_{N-1}}}
\frac{{n_{N-1}}_{S \mathbf{k_{N-1}}} {n_N}_{S \mathbf{k_N}}} 
{\big( \sum_{i = 1}^N \varepsilon_{n_i \mathbf{k_i}} \big) - E_N^\mathbf{q}}
\sum_{P}
{n_{i_{N-1}}^*}_{S' \mathbf{k_{i_{N-1}}}} 
{n_{i_N}^*}_{S' \mathbf{k_{i_N}}}
\gamma_{n_{i_1} \cdots n_{i_{N-2}} S'}^{\mathbf{k_{i_1}} \cdots \mathbf{k_{i_{N-2}}}} (\mathbf{q})
\label{eqn:gammaN},
\end{align}
\end{widetext}
where we define a renormalized parameter set
$
\gamma_{n_1 \cdots n_{N-2} S}^{\mathbf{k_1} \cdots \mathbf{k_{N-2}}} (\mathbf{q})
= \sum_{n_{N-1} n_N \mathbf{k_{N-1}}}
\alpha_{n_1 \cdots n_{N-1} n_N}^{\mathbf{k_1} \cdots \mathbf{k_{N-1}}} (\mathbf{q})
{n_{N-1}}_{S \mathbf{k_{N-1}}} {n_N}_{S \mathbf{k_N}}
$
for convenience. We emphasize that Eq.~(\ref{eqn:gammaN}) is formally exact for any 
$N \ge 2$, and it is one of our central results in this work.

For a given $N$, it is possible to reduce Eq.~(\ref{eqn:gammaN}) to a much simpler 
form by making extensive use of the relation
$
\gamma_{n_1 \cdots n_i \cdots n_j \cdots n_{N-2} S}^{\mathbf{k_1} \cdots \mathbf{k_i} 
\cdots \mathbf{k_j} \cdots \mathbf{k_{N-2}}} (\mathbf{q})
= 
\gamma_{n_1 \cdots n_j \cdots n_i \cdots n_{N-2} S}^{\mathbf{k_1} \cdots \mathbf{k_j} 
\cdots \mathbf{k_i} \cdots \mathbf{k_{N-2}}} (\mathbf{q})
$
that follow from the exchange symmetry of identical bosons. 
For instance when $N = 2$, Eq.~(\ref{eqn:gammaN}) reduces to
\begin{align}
\gamma_{S} (\mathbf{q}) 
&= - \frac{U}{N_c} \sum_{n_1 n_2 S' \mathbf{k_1}}
\frac{{n_1}_{S \mathbf{k_1}} {n_2}_{S \mathbf{k_2}} {n_1^*}_{S' \mathbf{k_1}}{n_2^*}_{S' \mathbf{k_2}}}
{\varepsilon_{n_1 \mathbf{k_1}} + \varepsilon_{n_2 \mathbf{k_2}} - E_2^\mathbf{q}}
\gamma_{S'} (\mathbf{q}),
\label{eqn:gamma2}
\end{align}
where $\mathbf{k_2} = \mathbf{q - k_1}$. We note that this expression looks very 
similar (in fact formally identical) to that of the two-body bound states of
fermions~\cite{iskin21, orso22}. 
In addition it recovers the single-band result~\cite{javanainen10, sanders11} 
when the band and sublattice indices are dropped from the summation, 
and the Bloch factors set to unity in the numerator. 
Similarly when $N = 3$, Eq.~(\ref{eqn:gammaN}) reduces to
\begin{align}
\label{eqn:gamma3}
&\gamma_{n_1 S}^{\mathbf{k_1}} (\mathbf{q}) 
= - \frac{U}{N_c} \sum_{n_2 n_3 S' \mathbf{k_2}}
\frac{{n_2}_{S \mathbf{k_2}} {n_3}_{S \mathbf{k_3}}}
{\big( \sum_{i = 1}^3 \varepsilon_{n_i \mathbf{k_i}} \big) - E_3^\mathbf{q}}
\\
&\times \big[
{n_2^*}_{S' \mathbf{k_2}}{n_3^*}_{S' \mathbf{k_3}}
\gamma_{n_1 S'}^{\mathbf{k_1}} (\mathbf{q}) 
+
{n_1^*}_{S' \mathbf{k_1}}{n_3^*}_{S' \mathbf{k_3}}
\gamma_{n_2 S'}^{\mathbf{k_2}} (\mathbf{q}) 
\nonumber \\
& \qquad \qquad \qquad \qquad \qquad \quad
+ 
{n_1^*}_{S' \mathbf{k_1}} {n_2^*}_{S' \mathbf{k_2}}
\gamma_{n_3 S'}^{\mathbf{k_3}} (\mathbf{q}) 
\big],
\nonumber
\end{align}
where $\mathbf{k_3} = \mathbf{q - k_1 - k_2}$. Upon a change of the summation variable 
$\mathbf{k_2}$, it can be shown that the second and third terms in the square 
bracket have exactly the same contribution after the summations. 
Thus Eq.~(\ref{eqn:gamma3}) can be equivalently written as
\begin{align}
\label{eqn:gamma3m}
&\gamma_{n_1 S}^{\mathbf{k_1}} (\mathbf{q}) 
= - \frac{U}{N_c} \sum_{n_2 n_3 S' \mathbf{k_2}}
\frac{{n_2}_{S \mathbf{k_2}} {n_3}_{S \mathbf{k_3}}}
{\big( \sum_{i = 1}^3 \varepsilon_{n_i \mathbf{k_i}} \big) - E_3^\mathbf{q}}
\\
&\times \big[
{n_2^*}_{S' \mathbf{k_2}}{n_3^*}_{S' \mathbf{k_3}}
\gamma_{n_1 S'}^{\mathbf{k_1}} (\mathbf{q}) 
+
2{n_1^*}_{S' \mathbf{k_1}}{n_3^*}_{S' \mathbf{k_3}}
\gamma_{n_2 S'}^{\mathbf{k_2}} (\mathbf{q}) 
\big].
\nonumber
\end{align}
Note that this expression also recovers the single-band result~\cite{mattis86, valiente10} 
when the band and sublattice indices are dropped from the summation, and the Bloch 
factors set to unity. Furthermore when $N = 4$, Eq.~(\ref{eqn:gammaN}) reduces to
\begin{align}
\label{eqn:gamma4}
&\gamma_{n_1 n_2 S}^{\mathbf{k_1} \mathbf{k_2}} (\mathbf{q}) 
= - \frac{U}{N_c} \sum_{n_3 n_4 S' \mathbf{k_3}}
\frac{{n_3}_{S \mathbf{k_3}} {n_4}_{S \mathbf{k_4}}}
{\big( \sum_{i = 1}^4 \varepsilon_{n_i \mathbf{k_i}} \big) - E_4^\mathbf{q}}
\\
 & \times \big[
{n_3^*}_{S' \mathbf{k_3}}{n_4^*}_{S' \mathbf{k_4}}
\gamma_{n_1 n_2 S'}^{\mathbf{k_1} \mathbf{k_2}} (\mathbf{q}) 
+
{n_2^*}_{S' \mathbf{k_2}}{n_4^*}_{S' \mathbf{k_4}}
\gamma_{n_1 n_3 S'}^{\mathbf{k_1} \mathbf{k_3}} (\mathbf{q}) 
\nonumber \\
&+
{n_2^*}_{S' \mathbf{k_2}}{n_3^*}_{S' \mathbf{k_3}}
\gamma_{n_1 n_4 S'}^{\mathbf{k_1} \mathbf{k_4}} (\mathbf{q}) 
+
{n_1^*}_{S' \mathbf{k_1}}{n_4^*}_{S' \mathbf{k_4}}
\gamma_{n_2 n_3 S'}^{\mathbf{k_2} \mathbf{k_3}} (\mathbf{q}) 
\nonumber \\
&+
{n_1^*}_{S' \mathbf{k_1}}{n_3^*}_{S' \mathbf{k_3}}
\gamma_{n_2 n_4 S'}^{\mathbf{k_2} \mathbf{k_4}} (\mathbf{q}) 
+
{n_1^*}_{S' \mathbf{k_1}}{n_2^*}_{S' \mathbf{k_2}}
\gamma_{n_3 n_4 S'}^{\mathbf{k_3} \mathbf{k_4}} (\mathbf{q}) 
\big],
\nonumber
\end{align}
where $\mathbf{k_4} = \mathbf{q - k_1 - k_2 - k_3}$.
We again note that this expression recovers the single-band result~\cite{kornilovitch22} 
when the band and sublattice indices are dropped from the summation, and the Bloch 
factors set to unity. 
Moreover, upon a change of the summation variable $\mathbf{k_3}$, it can be 
shown that the second and third terms as well as the fourth and fifth terms in 
the square bracket have exactly the same contributions after the summations. 
Thus Eq.~(\ref{eqn:gamma4}) can be equivalently written as
\begin{align}
\label{eqn:gamma44}
&\gamma_{n_1 n_2 S}^{\mathbf{k_1} \mathbf{k_2}} (\mathbf{q}) 
= - \frac{U}{N_c} \sum_{n_3 n_4 S' \mathbf{k_3}}
\frac{{n_3}_{S \mathbf{k_3}} {n_4}_{S \mathbf{k_4}}}
{\big( \sum_{i = 1}^4 \varepsilon_{n_i \mathbf{k_i}} \big) - E_4^\mathbf{q}}
\\
 & \times \big[
{n_3^*}_{S' \mathbf{k_3}}{n_4^*}_{S' \mathbf{k_4}}
\gamma_{n_1 n_2 S'}^{\mathbf{k_1} \mathbf{k_2}} (\mathbf{q}) 
+
2 {n_2^*}_{S' \mathbf{k_2}}{n_4^*}_{S' \mathbf{k_4}}
\gamma_{n_1 n_3 S'}^{\mathbf{k_1} \mathbf{k_3}} (\mathbf{q}) 
\nonumber \\
&+
2 {n_1^*}_{S' \mathbf{k_1}}{n_4^*}_{S' \mathbf{k_4}}
\gamma_{n_2 n_3 S'}^{\mathbf{k_2} \mathbf{k_3}} (\mathbf{q}) 
+
{n_1^*}_{S' \mathbf{k_1}}{n_2^*}_{S' \mathbf{k_2}}
\gamma_{n_3 n_4 S'}^{\mathbf{k_3} \mathbf{k_4}} (\mathbf{q}) 
\big].
\nonumber
\end{align}
By following a similar strategy, Eq.~(\ref{eqn:gammaN}) can be used to deduce the 
relevant integral equations for all other multimers with $N \ge 5$. 
As a numerical illustration, next we apply our theory to a one-dimensional two-band 
model, and benchmark its results with the DMRG simulations.

\section{Numerical Illustration}
\label{sec:ni}

Our theoretical analysis given above is valid for both attractive ($U < 0$) and 
repulsive ($U > 0$) interactions. However, given the symmetry of the Bloch bands, 
i.e., see below for $\varepsilon_{s k} (t,t') = -\varepsilon_{s k} (-t, -t')$, 
and that of the Bose-Hubbard Hamiltonian, 
i.e., $H(t,t', U) = - H(-t,-t',-U)$, 
below we consider only $U < 0$, 
since we are typically interested in the lowest-energy bound states that are 
below the continuum dissociation thresholds. That is our lowest-energy bound states for 
the $U < 0$ case correspond to the highest-energy bound-states above the continuum 
thresholds when $U > 0$.

\subsection{Sawtooth Lattice}
\label{sec:sl}

\begin{figure}[!htb]
    \centering
    \includegraphics[width=0.9\columnwidth]{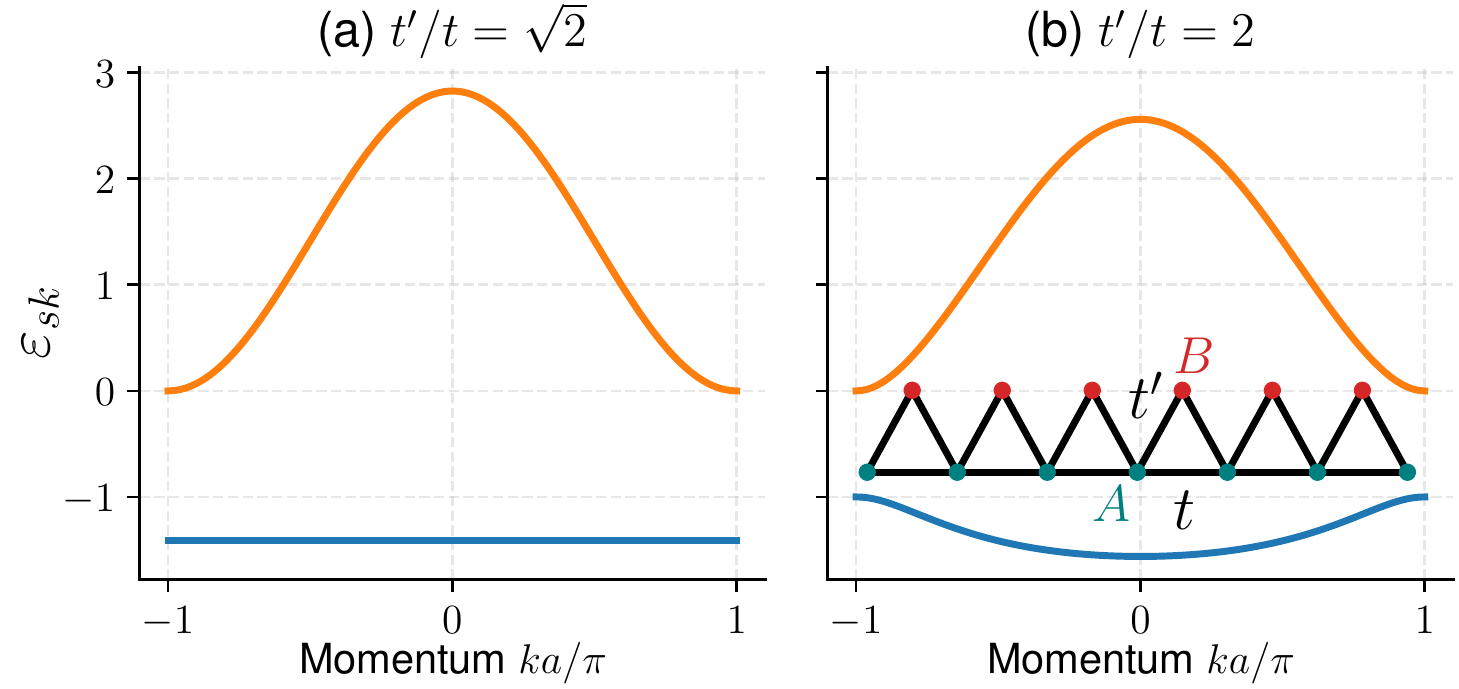}
    \caption{
    Bloch bands $\varepsilon_{s k}$ in a sawtooth lattice in the first BZ 
    for (a) $t = t'/\sqrt{2}$ and (b) $t = t'/2$. 
    Hopping parameters are also sketched in (b).
    Lower band is completely flat in (a) where 
    $\varepsilon_{-,k} = -\sqrt{2}t' \approx -1.414t'$.
    Ground-state energy is $(1-\sqrt{17})t'/2 \approx -1.561t'$ in (b).
    }
    \label{fig:1body}
\end{figure}

For simplicity here we choose a sawtooth lattice that features two Bloch bands in the 
first BZ (say $s = \{+, -\}$ bands) due to its $N_b = 2$ sublattice sites in a unit cell 
(say $S = \{A, B\}$ sublattices). We allow hopping between nearest-neighbor sites only, 
and set 
$
t_{Aj;Ai} = - t
$
with $j = i \pm 1$ and $t \ge 0$,
$
t_{Bj;Bi} = 0
$
and 
$
t_{Bi;Ai} = t_{Bj;Ai} = - t'
$
with $j = i-1$ and $t' \ge 0$. These are sketched in Fig.~\ref{fig:1body}(b).
The non-interacting Hamiltonian can be written as
$
\mathcal{H}_0 = \sum_k \psi_{k}^\dagger
\big(d_k^0 \sigma_0 + \mathbf{d}_k \cdot \boldsymbol{\sigma} \big)
\psi_{k},
$
where
$
\psi_{k} = ( c_{A k} \,\, c_{B k} )^\mathrm{T}
$
is a sublattice spinor, $-\pi/a < k \le \pi/a$ is in the first BZ with $a$ the lattice spacing, 
$
d_k^0 = t \cos(k a),
$
$\sigma_0$ is a $2\times2$ identity matrix,
$
\mathbf{d}_k = (d_k^x, d_k^y, d_k^z)
$
is a field vector with elements
$
d_k^x = t' + t' \cos(k a),
$
$
d_k^y = t' \sin(k a)
$
and 
$
d_k^z = t \cos(k a),
$
and
$
\boldsymbol{\sigma}  = (\sigma_x, \sigma_y, \sigma_z)
$
is a vector of Pauli spin matrices. The dispersion of the Bloch bands and the sublattice 
projections of the corresponding Bloch states can be written as
\begin{align}
\varepsilon_{s k} &=  d_k^0 + s d_k, \\
s_{A k} &= \frac{- d_k^x + id_k^y}{\sqrt{2d_k(d_k - sd_k^z)}}, \\
s_{B k} &= \frac{d_k^z - sd_k}{\sqrt{2d_k(d_k - sd_k^z)}},
\end{align}
where the index $s = \pm$ denotes the upper and lower bands, respectively, 
and $d_k$ is the magnitude of $\mathbf{d}_k$. Note that the lower band 
$\varepsilon_{-, k} = -2t$ is flat and dispersionless when $t'/t = \sqrt{2}$. 
This is shown in Fig.~\ref{fig:1body}(a). In addition one can recover the 
usual linear-chain model~\cite{valiente10} by setting $t = 0$, however with the 
caveat that its single cosine-band appears as two bands in our BZ. 
This is because the usual BZ $(-\pi/a_u, \pi/a_u]$ is folded into half 
$(-\pi/a, \pi/a]$ for our lattice spacing $a = 2a_u$,
leading to two bands that are symmetric around zero energy and no band gap 
in between. 

\subsection{Two-body spectrum}
\label{sec:twbs}

\begin{figure*}[!htb]
    \centering
    \includegraphics[width=1.99\columnwidth]{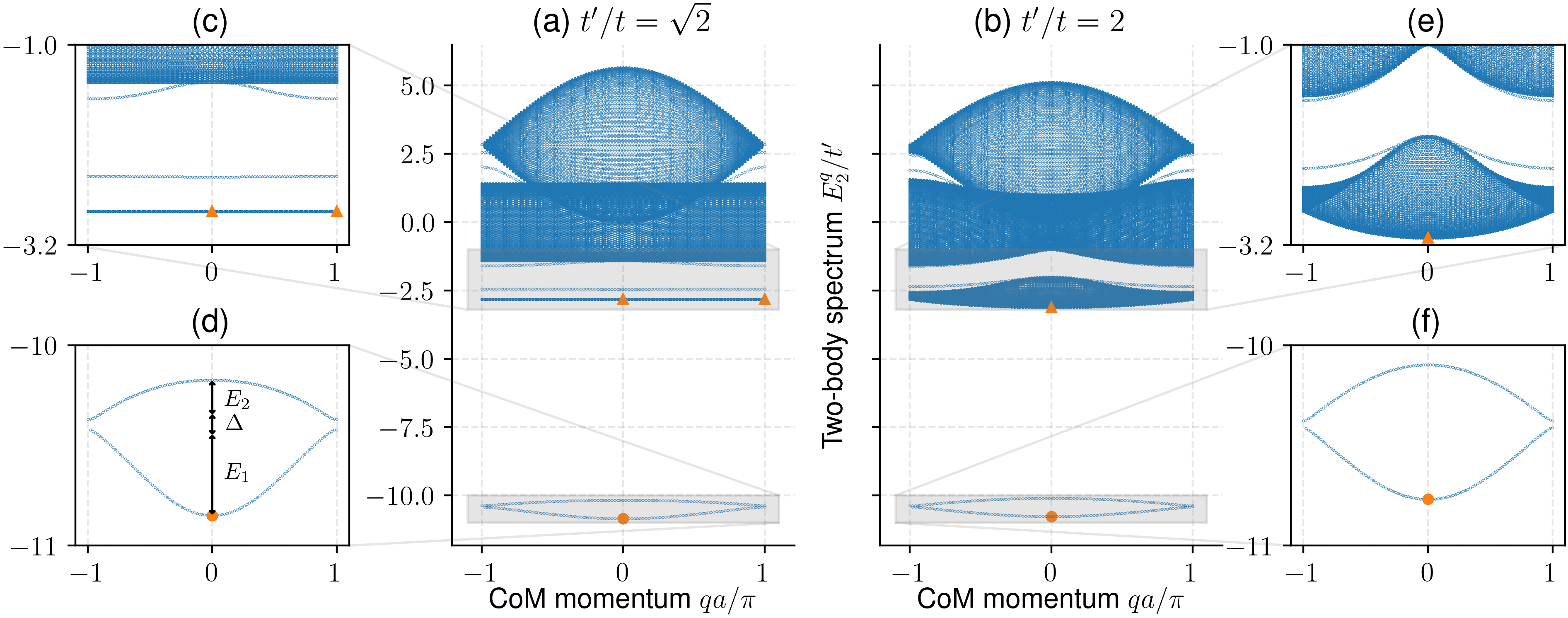}
    \caption{Two-body spectum $E_2^q$ of the attractive ($U=-10t'$) Bose-Hubbard model in a 
    sawtooth lattice for (a) $t = t'/\sqrt{2}$ and (b) $t = t'/2$.
    Insets (d) and (f) are zooms to the lower and upper onsite dimer branches.
    Their bandwidths $E_1$ and $E_2$, and the band gap $\Delta$ in between are 
    shown in Fig.~\ref{fig:2bodybw} as a function of $U/t'$. 
    Insets (c) and (e) are zooms to the monomer-monomer continua (i) and (ii),
    and two distinct offsite dimer branches in between. In (c) all of the continuum 
    (i) states appear precisely at $-2\sqrt{2}t' \approx -2.83t'$, and the lower 
    offsite dimer branch is around $-2.45t'$. 
    Benchmarks with DMRG ground states for $\rho_{max} = 2$ and $\rho_{max} = 1$ 
    are shown, respectively, with orange-colored markers $\bullet$ and 
    \scalebox{0.8}{$\blacktriangle$}.
    } 
    \label{fig:2body} 
\end{figure*}

The simplest bound states are those of the two-boson dimers that are determined by 
Eq.~(\ref{eqn:gamma2}). However, since Eq.~(\ref{eqn:gamma2}) is formally identical to 
that of the two-fermion case, we skip their detailed analysis here and refer the 
reader to the recent literature~\cite{iskin22b, orso22}. 
Just like the fermion problem, it can be shown that, by representing Eq.~(\ref{eqn:gamma2}) as an $N_b \times N_b$ matrix equation for the $\gamma_S(\mathbf{q})$ parameters, 
there are $N_b$ distinct $E_2^\mathbf{q}$ bound states for a given $\mathbf{q}$. 
In the sawtooth lattice the lower and upper dimer branches are associated, 
respectively, with the symmetric and antisymmetric combinations of the underlying 
sublattice contributions to the two-body wave function but with different 
weights~\cite{iskin21, orso22}. 
The solutions are shown in Figs.~\ref{fig:2body}(d) and~\ref{fig:2body}(f), respectively, 
for $t = t'/\sqrt{2}$ and $t = t'/2$ as a function of $q$ when $U = -10t'$. 
Note that these dimer branches appear at the bottom of the two-body spectrum, 
and they are well-separated from the rest of the states in the energy band.

\begin{figure}[!htb]
    \centering
    \includegraphics[width=0.9\columnwidth]{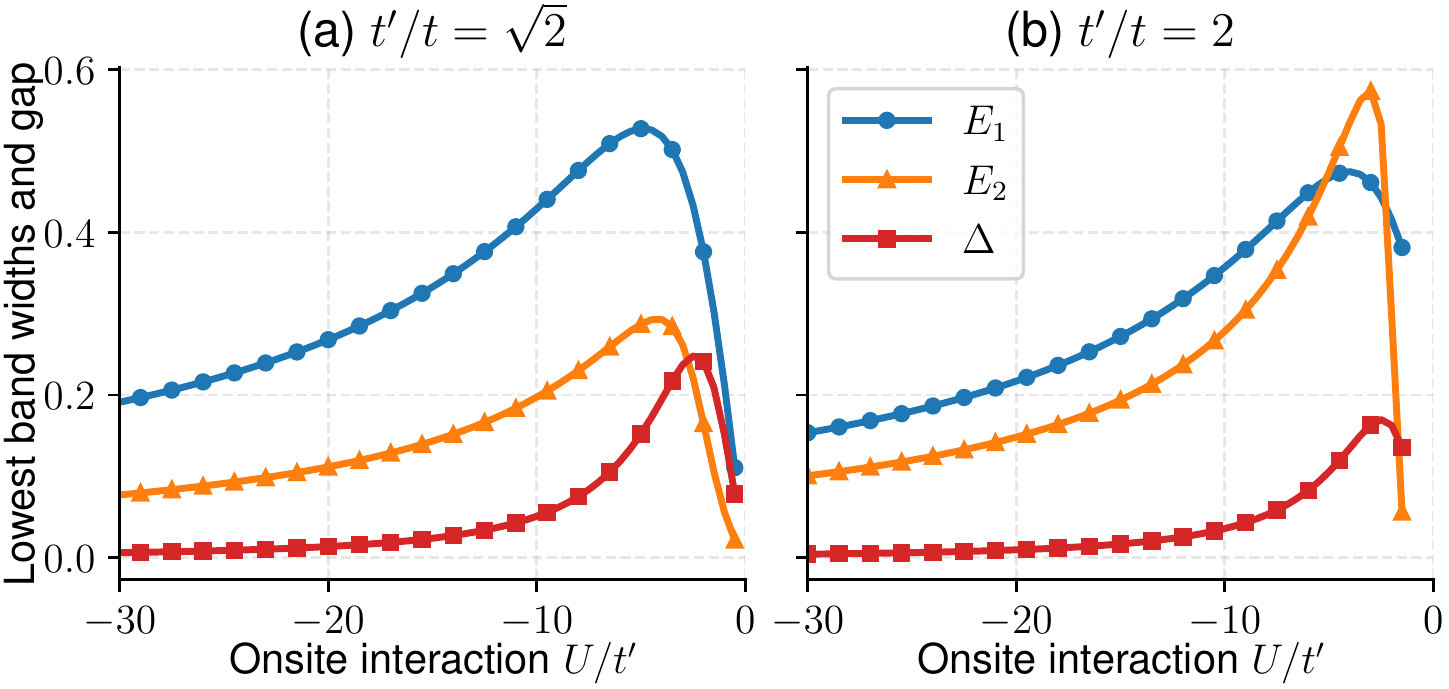}
    \caption{    
    Bandwidths $E_1$ and $E_2$, and band gap $\Delta$ of the onsite dimer branches 
    shown in Figs.~\ref{fig:2body}(d) and~\ref{fig:2body}(f). 
    These parameters play direct roles in the dimer-monomer continua shown in
    Figs.~\ref{fig:3body}(d) and~\ref{fig:3body}(f).
    }
    \label{fig:2bodybw}
\end{figure}

Starting with the usual single-band linear-chain lattice in the $t/t'\to 0$ limit 
which features a single dimer branch in its usual BZ~\cite{valiente08}, their 
doubling here in the sawtooth lattice can be traced back to doubling of the 
lattice spacing upon $t/t'\ne 0$ and folding of the usual BZ.
For instance in the flat-band case when $t = t'/\sqrt{2}$, the lower (upper) dimer 
branch has a bandwidth of $0.43t'$ ($0.20t'$), and its ground state $-10.85t'$ 
($-10.37t'$) is located at the origin (edge) $q = 0$ ($q = \pi/a$) of the BZ.
Likewise in the dispersive case when $t = t'/2$, the bandwidth $0.36t'$ ($0.28t'$) 
and ground state $-10.77t'$ ($-10.38t'$) are comparable. 
These are the so-called \textit{onsite} dimers since their binding energies depend 
strongly on $U$, approaching eventually to $U$ (with respect to the monomer-monomer 
continuum (i) discussed below) in the strong-coupling ($U/t' \to -\infty$) limit.
Note that there are two distinct onsite dimer branches in the sawtooth model 
since two bosons can be co-localized on one of the two sublattices. 
In addition the corresponding bandwidths ($E_1$ and $E_2$), and the band gap 
($\Delta$) of these dimer branches are shown in Fig.~\ref{fig:2bodybw} as a 
function of $U/t'$.
Altogether these results clearly show that the presence of a flat band does 
not make much impact on the onsite dimers, i.e., they have a sizeable dispersion 
at finite $U$ in general. This is because the introduction of an additional 
infinitely-massive flat-band monomer reduces the effective band mass of the 
resultant dimers from its bare value (which is infinite at $U= 0$) to a dressed one 
(which is finite at finite $U$) through interband processes~\cite{iskin21, iskin22, orso22}.
It is a somewhat counter-intuitive effect triggered by interactions in the
presence of multiple Bloch bands. The effective band mass of the onsite dimers 
is also known to exhibit a dip in the weak-coupling regime, and then it 
increases for stronger couplings leading to a more and more localized 
onsite dimer in space. This is because an onsite dimer can only move in the 
Bose-Hubbard model via virtual dissociation into monomers, and this reduces 
its motion by a factor of $U^{-1}$ (i.e., the dimer binding energy) 
from second-order perturbation theory.

Furthermore, in order to reveal the full two-body spectrum, here we recast 
Eq.~(\ref{eqn:gamma2}) as an eigenvalue problem in terms of the original 
variational parameters, i.e.,
\begin{align}
\label{eqn:gamma2SE}
& (\varepsilon_{n_1 \mathbf{k_1}} + \varepsilon_{n_2, \mathbf{q - k_1}} - E_2^\mathbf{q} )
\alpha_{n_1 n_2}^{\mathbf{k_1}} (\mathbf{q}) = 
\\
&- \frac{U}{N_c} \sum_{n_1' n_2' \mathbf{k} S}
{n_1^*}_{S \mathbf{k_1}} {n_2^*}_{S, \mathbf{q - k_1}} 
{n_2'}_{S, \mathbf{q-k}} {n_1'}_{S \mathbf{k}} 
\alpha_{n_1' n_2'}^{\mathbf{k}} (\mathbf{q}),
\nonumber 
\end{align}
and solve for $E_2^\mathbf{q}$. 
For a given $\mathbf{q}$, the solutions of Eq.~(\ref{eqn:gamma2SE}) require numerical 
diagonalization of a matrix whose size grows as $N_b^2 N_c$, and we typically choose 
$N_c = 100$ mesh points in the BZ in our two-body calculations. The results are 
shown in Figs.~\ref{fig:2body}(a) and~\ref{fig:2body}(b), respectively, 
for $t = t'/\sqrt{2}$ and $t = t'/2$ as a function of $q$ when $U = -10t'$. 
We also checked that (not shown) our results reduce to that of the usual 
linear-chain model when $t = 0$~\cite{valiente08}.
In both figures it is easy to characterize the entire two-body continuum which 
has contributions from three distinct monomer-monomer continua. 
Starting with the lowest one in energy, these are identified by the occupation of
(i) two bosons in the lower Bloch band, 
(ii) one boson in the upper Bloch band and one boson in the lower Bloch band, and
(iii) two bosons in the upper Bloch band.
More importantly we find two additional dimer branches appearing in between the 
two consecutive continua. These are the so-called \textit{offsite} dimers since 
their binding energies depend weakly on $U$, leading eventually to a small constant 
in the $U/t' \to -\infty$ limit depending only on $t/t'$. 
These dimers consist of a monomer on one 
site and a monomer on another site even in the $U/t' \to -\infty$ limit, i.e., 
the closest the two monomers can be is on nearest-neighbor sites.
Such states do not appear in a single-band linear-chain model~\cite{valiente08} 
since there is only a single monomer-monomer continuum 
there~\footnote{
Since the lowest offsite dimer branch appears above the monomer-monomer
continuum (i) but not below it, we suspect that their peculiar binding mechanism 
is mediated by an effective nearest-neighbor repulsion that is induced by the 
interband processes. This can be revealed by, e.g., following Ref.~\cite{valiente10}, 
and deriving an effective hardcore extended Bose-Hubbard Hamiltonian for the 
offsite dimers in the strong-coupling limit through an adiabatic 
elimination of the onsite dimers. 
We believe theoretical modelling of the binding mechanisms for the various 
offsite dimers, offsite trimers, offsite tetramers, etc., are interesting 
research problems by themselves, and are beyond the scope of this paper. 
}. 
Their presence clearly shed light on some of the recent results~\cite{phillips15, mielke18}.
For instance in the flat-band case 
when $t = t'/\sqrt{2}$, while the continuum (i) is located precisely at 
$-2\sqrt{2}t'$, the offsite dimer branch above this continuum has a small bandwidth 
of $0.007t'$, and its ground state $-2.45t'$ is located at the origin $q = 0$ of the BZ.
These are illustrated in Fig.~\ref{fig:2body}(c). 
Note that the effective band mass of the lower offsite dimer branch is much larger in 
magnitude than those of the onsite dimer branches. 
The continuum (ii) states occupy a rectangular region in Fig.~\ref{fig:2body}(a) 
around zero energy because, for every monomer 
state in the upper Bloch band, there always exists a monomer state in the flat 
band whose total momenta add up to $q$, and this is possible for any given $q$. 
Thus the bandwidth of the continuum (ii) states is determined by the bandwidth of 
the upper Bloch band shown in Fig.~\ref{fig:2bodybw}(a).

In the next section, we show that the entire two-body spectrum, i.e., the two-body 
continua together with the onsite and offsite dimer branches, play equally important 
roles in the characterization of the three-body spectrum.

\subsection{Three-body spectrum}
\label{sec:thbs}

\begin{figure*}[!htb]
    \centering
    \includegraphics[width=1.99\columnwidth]{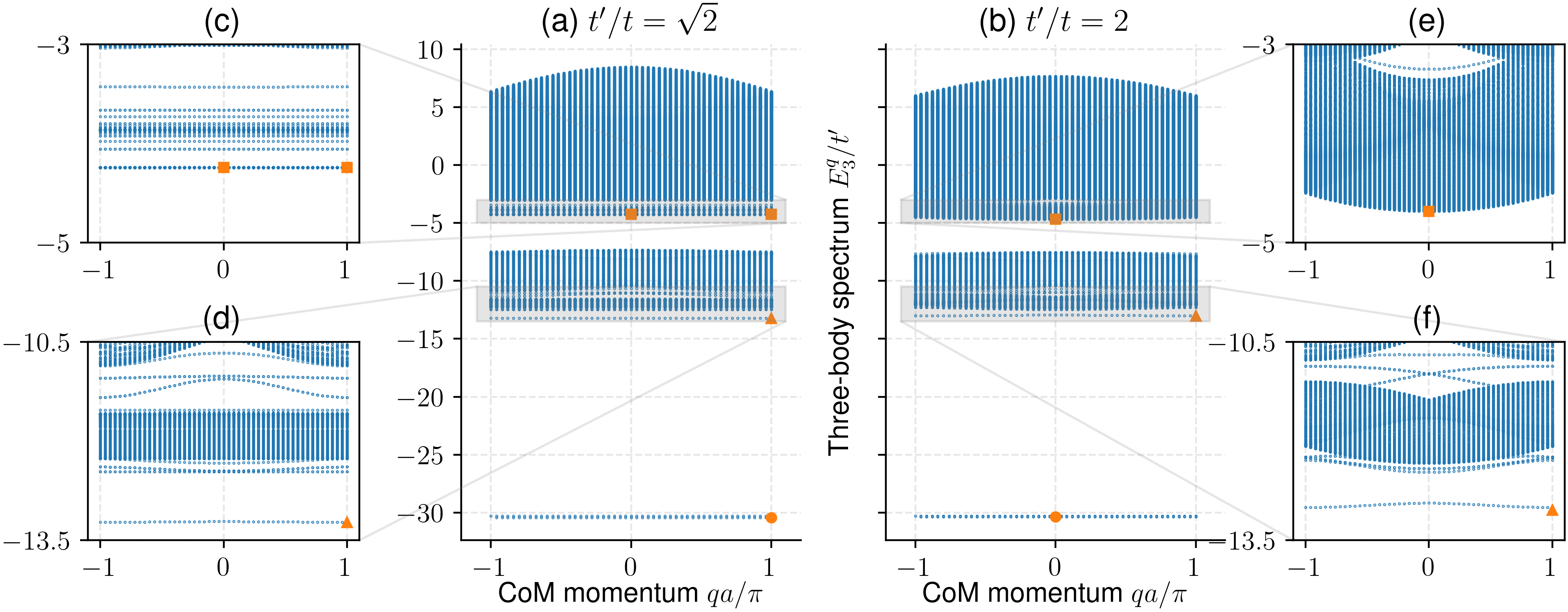} 
    \caption{Three-body spectum $E_3^q$ of the attractive ($U=-10t'$) Bose-Hubbard 
    model in a sawtooth lattice for (a) $t = t'/\sqrt{2}$ and (b) $t = t'/2$.
    Insets (d) and (f) are zooms to the offsite trimer branches nearby the 
    dimer-monomer continua (II-a), (II-b) and (II-c).
    Insets (c) and (e) are zooms to the monomer-monomer-monomer continua 
    (III-a) and (III-b), and the offsite trimer branches and offsite-dimer--monomer 
    continua in between. 
    In (c) all of the continuum (III-a) states appear precisely 
    at $-3\sqrt{2}t' \approx -4.24t'$, and the origin of the continuum of 
    states around $-2.45t' - \sqrt{2}t' \approx -3.86t'$ can be traced back to the lower 
    offsite dimer branch shown in Fig.~\ref{fig:2body}(c) around $-2.45t'$. 
    Benchmarks with the DMRG ground states for $\rho_{max} = 3$, $\rho_{max} = 2$ 
    and $\rho_{max} = 1$ are shown, respectively, with orange-colored markers $\bullet$,
    \scalebox{0.8}{$\blacktriangle$} and \scalebox{0.6}{$\blacksquare$}.
    } 
    \label{fig:3body} 
\end{figure*}

\begin{figure}[!htb]
    \centering
    \includegraphics[width=0.99\columnwidth]{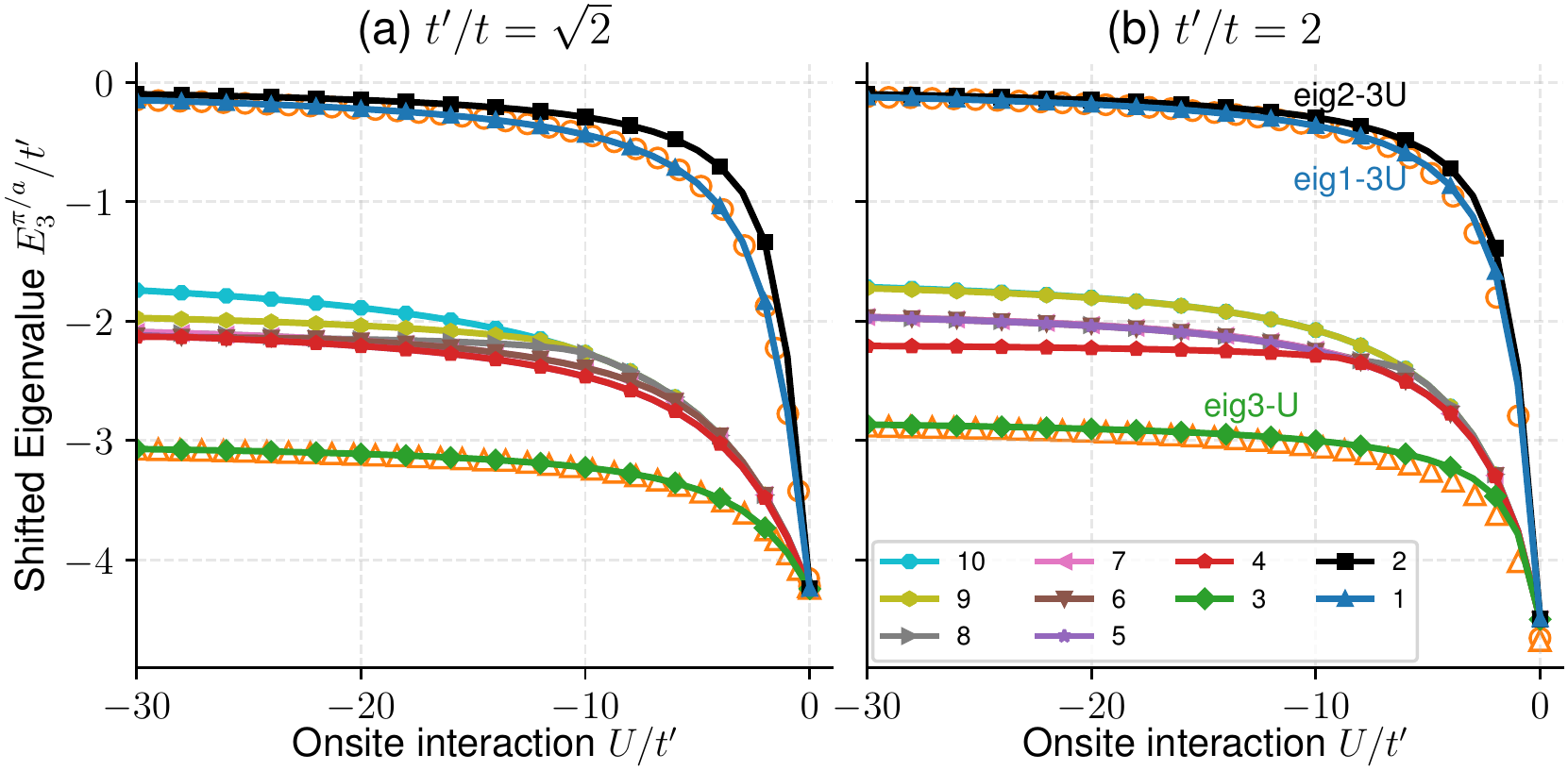}
    \caption{
    Lowest ten eigenvalues $E_3^{\pi/a}$ at $q = \pi/a$ for (a) $t = t'/\sqrt{2}$ 
    and (b) $t = t'/2$.
    First three of them belong, respectively, to the lower onsite trimer,
    upper onsite trimer and the lowest offsite trimer branches.
    Note that the latter branch emerges from the dimer-monomer continuum (II-a) 
    continuously (discontinuously) in the flat-band (dispersive-band) case 
    without (with) a threshold on $U$. In (b) this threshold is signalled 
    by the apparent degeneracy of the third eigenvalue with the rest 
    (fourth, fifth, sixth, etc.) at low $U/t'$.
    Onsite trimer energies are shifted by $3U$ but the rest are shifted by $U$
    for convenience.
    Benchmarks with the DMRG ground states for $\rho_{max} = 3$ and $\rho_{max} = 2$ 
    are shown, respectively, with orange-colored markers $\circ$ and 
    \scalebox{0.8}{$\triangle$}.
    } 
    \label{fig:eigs} 
\end{figure}

The three-body bound states can be determined by the integral Eq.~(\ref{eqn:gamma3}) 
through an iterative procedure~\cite{iskin22b}. Instead, in order to reveal the full 
three-body spectrum, where we recast Eq.~(\ref{eqn:gamma3}) as an eigenvalue problem 
in terms of the original variational parameters, i.e.,
\begin{align}
\label{eqn:gamma3SE}
&(\varepsilon_{n_1 \mathbf{k_1}} + \varepsilon_{n_2 \mathbf{k_2}} + \varepsilon_{n_3 \mathbf{k_3}} 
- E_3^\mathbf{q} )
\alpha_{n_1 n_2 n_3}^{\mathbf{k_1} \mathbf{k_2}} (\mathbf{q}) =
\\
&- \frac{U}{N_c} \sum_{n_2' n_3' \mathbf{k} S}
{n_2^*}_{S \mathbf{k_2}} {n_3^*}_{S \mathbf{k_3}}
{n_3'}_{S, \mathbf{q-k_1-k}} {n_2'}_{S \mathbf{k}}
\alpha_{n_1 n_2' n_3'}^{\mathbf{k_1} \mathbf{k}} (\mathbf{q}) 
\nonumber \\
&- \frac{U}{N_c} \sum_{n_1' n_3' \mathbf{k} S}
{n_1^*}_{S \mathbf{k_1}} {n_3^*}_{S \mathbf{k_3}}
{n_3'}_{S, \mathbf{q-k_2-k}} {n_1'}_{S \mathbf{k}}
\alpha_{n_1' n_2 n_3'}^{\mathbf{k} \mathbf{k_2}} (\mathbf{q}) 
\nonumber \\
&- \frac{U}{N_c} \sum_{n_1' n_2' \mathbf{k} S}
{n_1^*}_{S \mathbf{k_1}} {n_2^*}_{S \mathbf{k_2}} 
{n_2'}_{S, \mathbf{q-k_3-k}} {n_1'}_{S \mathbf{k}} 
\alpha_{n_1' n_2' n_3}^{\mathbf{k}, \mathbf{q-k_3-k}} (\mathbf{q}),
\nonumber 
\end{align}
and solve for $E_3^\mathbf{q}$. Recall that $\mathbf{k_3} = \mathbf{q - k_1 - k_2}$ 
when $N = 3$, and note that the exchange-symmetry constraints are imposed on 
the variational parameters by construction. The symmetry can be made 
explicit by properly doubling the terms on the right hand side (and dividing 
the entire expression by two), but we checked that the resultant equation and 
Eq.~(\ref{eqn:gamma3SE}) always produce identical solutions for the sawtooth model.
For a given $\mathbf{q}$, the solutions of Eq.~(\ref{eqn:gamma3SE}) 
require numerical diagonalization of large matrices whose size grows as 
$N_b^3 N_c^2$, and we typically choose $N_c = 50$ mesh points in the BZ in our 
three-body calculations. Decreasing it to $N_c = 40$ makes only minor changes.
The results are shown in Figs.~\ref{fig:3body}(a) and~\ref{fig:3body}(b), respectively, 
for $t = t'/\sqrt{2}$ and $t = t'/2$ as a function of $q$ when $U = -10t'$. 
We also checked that (not shown) our results reduce to that of the usual 
linear-chain model when $t = 0$~\cite{valiente10}.
In both figures the spectrum splits into three groups of states, 
i.e., starting with the lowest one in energy,:
(I) two trimer branches lie around $3U$ for a given $q$,
(II) dimer-monomer continua are packed around $U$, and
(III) monomer-monomer-monomer continua are packed around zero energy.
The group (I) branches are best resolved in the corresponding lowest-energy 
solutions (first ten of them) that are shown in Figs.~\ref{fig:eigs}(a)
and~\ref{fig:eigs}(b) as a function of $U/t'$ at $q = \pi/a$. 

First of all, similar to the two-body spectrum, we find that there are two 
trimer branches appearing at the bottom of the three-body spectrum, and they are 
well-separated from the rest of the states in the spectrum. 
Starting with the usual single-band linear-chain lattice in the $t/t'\to 0$ limit 
which features a single trimer branch in its usual BZ~\cite{valiente10}, 
their doubling here in the sawtooth lattice can again be traced back to doubling 
of the lattice spacing upon $t/t'\ne 0$ and folding of the usual BZ. 
These are the so-called \textit{onsite} trimers since their binding energies 
depend strongly on $U$, approaching eventually to $2U$ (with respect to the 
dimer-monomer continua) in the $U/t' \to -\infty$ limit~\cite{mattis86, valiente10}.
We again note that there are two distinct onsite trimer branches in the sawtooth 
model since three bosons can be co-localized on one of the two sublattices.
Unlike the highly dispersive onsite dimer branches, the onsite trimer branches 
have nearly-flat dispersions even in the weak-binding low-$U/t'$ regime,
i.e., since the effective band mass of the onsite trimers is much larger in magnitude 
than that of the onsite dimers, the onsite trimers are more localized in space.
This is because an onsite trimer can only move in the Bose-Hubbard model via 
virtual dissociation into monomers, and this reduces its motion by a factor 
of $U^{-2}$ from third-order perturbation theory.
For instance in the flat-band case when $t = t'/\sqrt{2}$, the lower (upper) 
onsite trimer branch has a small bandwidth of $0.011t'$ ($0.0047t'$), 
and its ground state $-30.4379t'$ ($-30.2911t'$) is located at the 
edge $q = \pi/a$ of the BZ. 
Their energy difference fits very well with $(\sqrt{2}t'/U) t'$ in the 
strong-coupling up to low $U/t' \sim -3$ values.
Likewise in the dispersive case when $t = t'/2$, the bandwidth $0.0046t'$ 
($0.0094t'$) is again small but the ground state $-30.3682t'$ ($-30.2939t'$) 
is located at the origin $q = 0$ (edge $q = \pi/a$). 

At the bottom of group (II) states, there are the so-called \textit{offsite} trimers 
since their binding energies depend weakly on $U$, leading eventually to 
a small constant (with respect to the dimer-monomer continua) 
in the $U/t' \to -\infty$ limit depending only on $t/t'$. 
These trimers consist of a dimer on one site and 
a monomer on another site even in the $U/t' \to -\infty$ limit, i.e., 
the closest the dimer and the monomer can be is on nearest-neighbor sites.
These states are shown in Figs.~\ref{fig:3body}(d) and~\ref{fig:3body}(f).
For instance in the flat-band case when $t = t'/\sqrt{2}$, 
the lowest offsite trimer branch has a small bandwidth of $0.0096t'$, 
and its ground state $-13.2256t'$ is located at the edge $q = \pi/a$ of the BZ. 
Likewise in the dispersive case when $t = t'/2$, its bandwidth $0.065t'$ is 
also small but its ground state $-13.0015t'$ is located at $q = \pi/a$. 
Thus the effective band mass of the lowest offsite trimers is also much larger 
in magnitude than that of the onsite dimers. 
Unlike the offsite dimers which appear only in the presence of multiple bands, 
these offsite trimers are known to appear also in a single-band linear-chain 
model but when $U$ is sufficiently strong. 
However, while there appears precisely a single trimer branch in the 
single-band model~\cite{valiente10}, here we observe several branches in a 
two-band model whose number depends strongly on 
$U/t'$~\footnote{
It turns out these offsite boson trimers are in many ways similar to the 
fermion trimers in the $(2+1)$-body problem~\cite{iskin22b}. For instance 
the fermion trimers are necessarily offsite and they are weakly-bound 
due to the Pauli exclusion principle preventing the formation of onsite trimers. 
What is astounding is that, in the case of sawtooth model, the low-energy 
spectrum of the $(2+1)$-body fermion problem coincides exactly (i.e., up to the 
machine precision) with excited states of the three-boson problem. 
Our variational calculations for the Hubbard and Bose-Hubbard models show that 
this is generally the case for any given set of $\{t, t', U\}$. 
It is such that the energy of the ground fermion trimer state coincides 
with the fifth lowest eigenvalue (i.e., third offsite trimer branch) of 
the boson one for any given CoM momentum $q$. 
In addition the energy of the excited fermion trimer state coincides 
with the seventh lowest eigenvalue (i.e., fifth offsite trimer branch) 
of the boson one. See Ref.~\cite{iskin22c} for a more detailed comparison.
}. 
Note that the gradual appearance of additional offsite trimer branches with 
increasing $U/t'$ is clearly seen in Fig.~\ref{fig:eigs}.
In addition the lowest trimer branch emerges from the dimer-monomer 
continuum continuously in the flat-band case without a threshold on $U$, 
which is signalled by the apparent nondegeneracy of the third eigenvalue 
for all $U/t'$ in Fig.~\ref{fig:eigs}(a).

Furthermore the group (II) states have contributions from four distinct 
dimer-monomer continua.
Starting with the lowest one in energy (which assumes $U/t'$ is sufficiently strong), 
these are identified by the occupation of
(II-a) two bosons in the lower onsite dimer branch and one boson in the lower Bloch band,
(II-b) two bosons in the upper onsite dimer branch and one boson in the lower Bloch band,
(II-c) two bosons in the lower onsite dimer branch and one boson in the upper Bloch band, and
(II-d) two bosons in the upper onsite dimer branch and one boson in the upper Bloch band.
For instance the first continuum (II-a) first appears around 
$-10.85t' - 1.41t' = -12.26t'$ and $-10.77t' - 1.56t' = -12.33t'$,
respectively, when $t = t'/\sqrt{2}$ and $t = t'/2$. 
These are shown in Figs.~\ref{fig:3body}(d) and~\ref{fig:3body}(f).
In the particular case when the lower Bloch band is flat, the bandwidths of 
continuum (II-a) and continuum (II-b) are determined, respectively, by solely the 
bandwidths ($E_1$ and $E_2$) of the lower and upper onsite dimer 
branches that are shown in Fig.~\ref{fig:2bodybw}(a). Note that the gap 
between continuum (II-a) and continuum (II-b) is determined by $\Delta$, 
and it is barely visible here.
In both figures we also find several additional offsite trimer branches in between 
the continuum (II-b) and continuum (II-c). There is also an additional offsite trimer 
branch above the continuum (II-d) which is not visible in the shown scale. 

Likewise the group (III) states have contributions from four distinct 
unbound monomer-monomer-monomer continua. 
Starting with the lowest one in energy, these are identified by the 
occupation of
(III-a) three bosons in the lower Bloch band,
(III-b) two boson in the lower Bloch band and one boson in the upper Bloch band,
(III-c) one boson in the lower Bloch band and two bosons in the upper Bloch band, and
(III-d) three bosons in the upper Bloch band.
For instance the first continuum (III-a) appears at 
$-3\sqrt{2}t' \approx -4.24t'$ and $3(1-\sqrt{17})t'/2 \approx -4.68t'$, 
respectively, when $t = t'/\sqrt{2}$ and $t = t'/2$. 
These are shown in Figs.~\ref{fig:3body}(c) and~\ref{fig:3body}(e).
In our numerical calculations, we did not find any offsite trimer branch
below the continuum (III-a). 
However, there can be offsite trimer branches in between continuum (III-a) 
and continuum (III-b). For instance these branches are clearly seen in 
the flat-band case when $t = t'/\sqrt{2}$. 
Right above them we also find a continuum of states that are packed around 
$-2.45t' - \sqrt{2}t' \approx -3.86t'$. 
These are shown in Fig.~\ref{fig:3body}(c).
It turns out the latter can be identified as an additional dimer-monomer 
continuum, emerging from the occupation of the lower offsite dimer branch 
that we found at $-2.45t'$ in Fig.~\ref{fig:2body}(c) and a monomer in the 
flat Bloch band. 
These states consist of three monomers that can at most be found on 
different nearest-neighbor sites even in the $U/t' \to -\infty$ limit.
Furthermore while the continuum (III-b) is expected to appear at 
$-2\sqrt{2}t'+ 0t' \approx -2.83t'$ in the flat-band case, we find a continuum of states 
starting around $-3.00t'$, which is barely visible at the top of Fig.~\ref{fig:3body}(c).
This is because there is yet another dimer-monomer continuum emerging from the 
occupation of the upper offsite dimer branch that we found right below the 
monomer-monomer continuum (iii) in Fig.~\ref{fig:2body}(c) and 
a monomer in the flat Bloch band. 

Having analyzed the two-body and three-body spectra for the sawtooth lattice 
via our exact variational results, next we benchmark them with the 
DMRG simulations.

\subsection{DMRG Simulations}
\label{sec:ds}
\begin{figure*}[!htb]
    \centering
    \includegraphics[width=1.86\columnwidth]{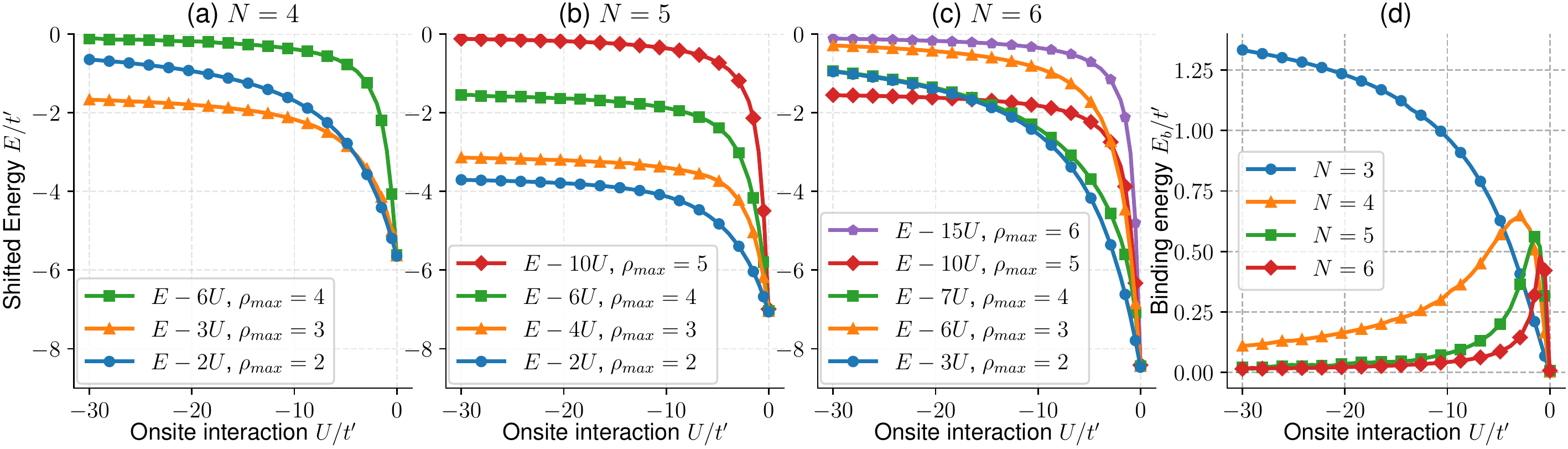}
    \caption{
    DMRG ground-state energy $E_N(\rho_{max})$ for (a) the tetramer ($N = 4$), 
    (b) pentamer ($N = 5$), and (c) hexamer ($N = 6$) states in a flat band 
    (i.e., $t = t'/\sqrt{2}$) when the local number of bosons is restricted to 
    $\rho_{max} = \{N, N-1, \cdots, 2\}$. 
    The energies are shifted by $E_{sat}$ for convenience. 
    Since the ground states for the dispersive case 
    (i.e., $t = t'/2$) are quantitatively very similar, they are not shown. 
    Binding energy $E_b$ for the lowest offsite $N$-body multimers are shown in (d).
    }
    \label{fig:Nbody} 
\end{figure*} 

Even though one can understand much of the two-body spectrum by looking at 
the Bloch bands alone, and in return can also keep track a large portion of 
the three-body spectrum by looking at the two-body spectrum, it is important 
to benchmark and verify our results with an independent calculation. 
Here we present our numerically-exact DMRG 
simulations~\cite{srwhite, schollwock11, itensor} on a large lattice with up to
$N_c = 100$ unit cells. 
Our checks include not only the ground states of the two-body and the three-body 
spectra (i.e., the ground states for the lower onsite dimer branches 
shown in Figs.~\ref{fig:2body}(d) and~\ref{fig:2body}(f), and for the lower onsite 
trimer branches shown in Figs.~\ref{fig:3body}(a) and~\ref{fig:3body}(b), 
respectively) but also the ground states for the offsite trimer branches that 
are shown in Figs.~\ref{fig:3body}(d) and~\ref{fig:3body}(f). 
While the former is achieved by allowing a large cutoff in the local Hilbert space
(i.e., local number of bosons) on each lattice site (i.e., any $\rho_{max} \ge 3$ 
is sufficient for the onsite trimers), the latter is achieved through restricting
$\rho_{max} = 2$ to that of an onsite dimer. Setting $\rho_{max} = 1$ determines 
the bottom edge of the continuum (III-a).

In the flat-band case when $t = t'/\sqrt{2}$, we note that the agreement 
between the variational approach and the DMRG simulations is almost perfect, 
i.e., the relative accuracy between the two is typically better than $0.1\%$ 
for all of the parameters that we considered.
Setting $\rho_{max} = 2$ works surprisingly well for the offsite trimers 
because the flat-band dimers are already in the strong-coupling limit 
when they first form at arbitrarily weak $U \ne 0$.
However, in the dispersive case when $t = t'/2$, while the agreement is again 
almost perfect for the onsite trimers, it is not as good for the offsite trimers 
when $U/t' \gtrsim -10$ is relatively weak. 
This shows that $\rho_{max} = 2$ provides only a qualitatively accurate 
description (but not a quantitative one) for the offsite trimers in the 
weak-coupling regime. 
Thus our successful benchmark for the strong-coupling $U/t' \to -\infty$ limit 
suggests that the states in the lowest offsite trimer branch can really be 
thought of a bound state between an onsite dimer and a monomer occupying 
different lattice sites, and hence the origin of their name \textit{offsite}. 
Note that a similar strategy does not work for 
the ground states of the offsite dimer branches that are shown in 
Figs.~\ref{fig:2body}(c) and~\ref{fig:2body}(e), 
since they appear above the monomer-monomer continuum (i) but not below it, 
i.e., setting $\rho_{max} = 1$ determines the bottom edge of the continuum (i).

Given our almost perfect benchmark for the ground states of onsite dimers, 
onsite trimers and offsite trimers, we also perform DMRG simulations for the
onsite tetramers, offsite tetramers, onsite pentamers, offsite pentamers, etc. 
That is we study the ground states of various $N$-body multimers by setting 
$\rho_{max} = \{N, N-1, ..., 2\}$ for a given $N$. 
These results are shown in Fig.~\ref{fig:Nbody}. 
For a given $N = \{4,5,6\}$ and $\rho_{max}$, the DMRG simulation gives the ground 
state energy for the multimer that has at most $\rho_{max}$ particles on a given site.
As shown in Fig.~\ref{fig:Nbody}, the ground-state energy $E_N(\rho_{max})$ of 
the corresponding multimer generally approaches to
\begin{align}
E_{sat} = n \frac{\rho_{max}(\rho_{max}-1)}{2} U + \frac{m(m-1)}{2} U
\end{align} 
in the strong-coupling limit, where the integers $n \ge 0$ and $0 \le m < \rho_{max}$ 
are such that
$
N = n \rho_{max} + m.
$
This is because, for a given $\rho_{max}$, the ground state of the offsite 
$N$-body multimer consists of $n$ onsite $\rho_{max}$-body multimers 
and an onsite $m$-body multimer that are all on different sites.
For instance while the $\rho_{max} = N$ case recovers the expected ground state 
of the onsite $N$-body multimer that is strongly localized on a single lattice
site~\cite{mattis86}, the $\rho_{max} = N - 1$ case gives the ground state 
of the offsite $N$-body multimer that consists of at most an onsite $(N-1)$-body 
multimer on one site and a monomer on another site. 
Setting $\rho_{max} = 1$ generally determines the bottom edge of the continuum 
that is characterized by $N$ unbound (non-interacting) monomers in the 
lowest Bloch band. 

Our numerical results confirm that the binding energy of the onsite $N$-body 
multimers increases with $N$ and $U$ without a bound, i.e., it trivially goes 
as $E_b = -E_N(N) + E_{N-1}(N-1) + E_1(1) \approx -(N-1)U$ in the strong-coupling 
limit. Therefore, similar to the onsite dimers and onsite trimers, we expect 
the effective band mass of all onsite $N$-body multimers to increase in the 
strong-coupling regime, where these states become more and more localized 
in space. In addition the larger the size $N$ of the onsite multimer, 
the higher its effective band mass is for a given $U$.

More importantly our results also suggest that the binding energy of the 
lowest offsite $N$-body multimers can generally be defined as
\begin{align}
E_b = -E_N(N-1) + E_{N-1}(N-1) + E_1(1)
\label{eqn:Eb}
\end{align}
in the strong-coupling limit. Furthermore this definition is expected
to be valid for all $N$ and $U$ in the flat-band case, since the onsite 
multimers are already in the strong-coupling limit when they first form 
at arbitrarily weak $U \ne 0$. In fact we checked this for low-$N$ 
values by considering all other possible dissociation processes, 
and verified that Eq.~(\ref{eqn:Eb}) gives the largest binding energy. 
For instance $E_b$ corresponds precisely to the energy gap between the 
lowest offsite trimer branch and the dimer-monomer continuum (II-a) 
that are shown in Fig.~\ref{fig:3body}(d).
Our DMRG results are shown in Fig.~\ref{fig:Nbody}(d) as a function 
of $U/t'$ for $N = \{3,4,5,6\}$. 
For the offsite trimers with $N = 3$, $E_b$ increases monotonously 
from $0$ to a constant value in the strong-coupling limit, 
signalling that the onsite dimer and the monomer are eventually on the 
nearest-neighbor sites. Similar to the usual linear-chain 
model~\cite{valiente10}, their binding mechanism is expected to be an 
effective particle-exchange interaction which depends only on the 
hopping parameters $t/t'$ but not on $U$. This is because 
the exchange process $|2, 1 \rangle \to |1, 2 \rangle$
does not involve $U$.
On the other hand, for larger offsite multimers with $N \ge 4$, 
$E_b$ first exhibits a peak in the weak-coupling regime and then it decays 
for stronger couplings. This indicates that the effective particle-exchange 
interaction between the constituents of the offsite $N$-body multimer, 
i.e., the $|N-1, 1 \rangle \to |N-2, 2 \rangle$ process between the 
onsite $(N-1)$-body multimer and the monomer, not only depends on $U$ 
but also decreases in the strong-coupling regime. 
This is because, given that increasing the coupling strength 
strongly localizes the onsite $(N-1)$-body multimer state in space 
(i.e., increases both of its binding energy and effective band mass), it becomes 
energetically more difficult to exchange one of its constituents with the 
monomer on another site due to decreasing overlap of their wave functions. 
This also explains why $E_b$ decays faster for larger $N$.

\section{Conclusion}
\label{sec:conc}

In summary here we derived exact integral equations for the dimers, trimers, tetramers, 
and other multimers that are described by a multiband Bose-Hubbard model with generic 
hoppings and an onsite attractive interaction. As an illustration, we calculated the 
two-body and three-body spectra in a sawtooth model, and revealed the presence of both 
the weakly-bound offsite dimer states which consist of two monomers on different sites 
even in the strong-coupling ($U/t' \to -\infty$) limit, and the weakly-bound offsite 
trimer states which consist of either a dimer on one site and a monomer on another site
or three monomers on three different sites. 
We benchmarked the ground states of onsite dimers, onsite trimers and offsite 
trimers with the DMRG simulations, and presented additional DMRG results 
for the ground states of onsite tetramers, offsite tetramers, onsite pentamers, 
offsite pentamers and for those of other multimers. 

Even though we restricted our numerical analysis here to a one-dimensional 
model, i.e., due mainly to our benchmarking capacity with the DMRG simulations, 
our variational results may find practical applications in future few-body studies, 
since they are readily applicable to all sorts of lattices in all dimensions.
In particular while the Efimov effect is known to be absent in one or two 
dimensions, it may be studied in three-dimensional crystals~\cite{mattis86}.
As an outlook it may be useful to develop a simpler effective model that 
reveals the binding mechanism of offsite $N$-body multimers in the presence of 
multiple bands. It is expected to be very similar to that of the offsite trimers 
in a single-band lattice~\cite{valiente10}, i.e., there must be an effective 
particle-exchange interaction between an onsite $(N-1)$-body multimer on one 
site and a monomer on another site, between an onsite $(N-2)$-body multimer 
on one site and a dimer on another site, between an onsite $(N-2)$-body 
multimer on one site and two monomers on two other sites, etc.
In addition one can easily extend our approach to study bound-state formation 
in multiband Bose-Fermi mixtures.

\begin{acknowledgments}
We thank M. Valiente for his comments and suggestions. 
A. K. is supported by  T{\"U}B{\.I}TAK 2236 Co-funded Brain Circulation Scheme 2 
(CoCirculation2) Project No. 120C066.
\end{acknowledgments}

\bibliography{refs}

\end{document}